\documentstyle{l-aa}
\input epsf.sty
\def\eps{\varepsilon}
\def\o{\omega}
\def\O{\Omega}
\def\a{\alpha}

\def\d{\delta}

\def\e{\eta}

\def\g{\gamma}

\thesaurus{04(11.01.2; 11.10.1; 11.14.1)}

\begin{document}
\headnote{Letter to the Editor}

\title{On the nonthermal emission in active galactic nuclei}

\author{R. T. Gangadhara$^1$ and H. Lesch$^2$}
\institute{$^1$Max--Planck--Institut f\"ur Radioastronomie,
Auf dem H\"ugel 69, D--53121 Bonn, Germany\\
$^2$Institut f\"ur Astronomie und Astrophysik der
Universit\"at M\"unchen,
Scheinerstra$\beta$e 1, 81679 M\"unchen, Germany}
\date{Received February 2, 1997; accepted May 14, 1997 }

\maketitle

\begin{abstract}

     We consider the role of centrifugal force on the energetics of electrons moving along
the magnetic field lines of spinning active galactic nuclei. We find the energy gained by
charged particles
against inverse Compton scattering and/ synchrotron radiation losses, become quite significant in
the region close to the light cylinder. The particles accelerated by the centrifugal force become a
part of the jet material. The scattering of UV--photons against the energetic electrons lead to the
generation of X--ray and $\gamma$-ray photons within the light cylinder, and outside, they radiate the
non--thermal optical radiation via the synchrotron emission.

\keywords{ Galaxis: active -- nuclei -- jets}

\end{abstract}

\section{Introduction}

One of the most challenging problems in the area of active galactic nuclei
(AGN) is the origin of nonthermal continuum emission (Ekers et al. 1996 and 
references therein).
In the radio-loud AGN, the central engine has sufficient angular
momentum and rotation energy to maintain jets and the non-thermal radiation.
In-situ acceleration within jets is required to explain the
existence of highly relativistic particles at radii where the light
travel time from the nucleus is orders of magnitude larger than the
radiative lifetime of the particles. In this paper, we consider the
effect of AGN rotation on the charged particles moving along magnetic
field lines. Both relativistic electrons and large scale magnetic
fields are clearly shown to exist in AGN (e.g., Lovelace \& Contopoulos~1991).
The polarized continuum emission
of the very centre directs towards a particle population at least at GeV level. On the other
hand it has been shown that magnetic fields are involved in the driving
mechanisms and collimation of jets (Camenzind 1996). The problem is, however, to
couple the magnetohydrodynamical models with the observed relativistic
particle population. It is the aim of our contribution to show that within
a rotating magnetosphere which is at the foot points of a jet there exists
a natural source for the relativistic particles. Injecting particles along the
field lines and following them until the light cylinder (where the rotation
speed reaches the velocity of light) we show that despite the intense
UV-radiation field of the accretion disk, particles can reach high energies.

It is the aim of
our contribution to present a possible connection of the  MHD-scenarios
for the jet production like the models developed by Camenzind (1996) and
the acceleration of particles and accompanying radiation losses.
In the next section we consider the acceleration of particles in
a rotating magnetosphere. Section 3 contains the influence of inverse
Compton scattering and synchrotron radiation on the particle
acceleration. Finally, we present some conclusions.

\section{Centrifugal acceleration of charged particles moving along rotating
     magnetic field lines}

 The idea that the magnetized winds emanating from rotating objects can
extract angular momentum and energy, has been dealt in the variety of
contexts, for example, the production of MHD-winds in AGN (Camenzind 1986),
break down of solar rotation (Weber \& Davis 1967; Mestel 1968; Michel
1969), and pulsar wind production (Goldreich \& Julian 1970; Kennel et al.
1983; Gangadhara 1995, 1996). In what follows, we explore the possibility that
rotational energy can be extracted from a rapidly rotating supermassive
object by the magnetized plasma particles.

The polarization properties and spectra of AGN indicate
the synchrotron radiation from relativistic electrons moving in the
magnetic field at 1 parsec $\sim 1$G. There are
substantial evidences both from the theory as well as observations about the fact
that magnetic fields play an important role in the production and
collimation of jets (e.g., Lesch et al. 1989; Camenzind 1996;
Blandford 1990; Wiita 1993).

 Consider a typical galactic nucleus associated with the large scale magnetic field
lines crossing the light cylinder. The radius of light cylinder
in the case of AGN is $r_{\rm LC}=c/\O,$ where $\O$ is the
angular velocity of the field lines. If the field lines are rotating
then the particles gain rotation energy when they move from slowly
rotating region (central engine) to the fast rotating region (light cylinder).
The equation of motion of a particle in the rotating frame is given
by (Gangadhara 1996)
\begin{equation}
\frac{d}{dt}\left(m\frac{d r}{dt}\right)\hat{e}_{\rm r}=\vec F_{\rm B}+
\vec F_{\rm c}+\vec F_{\rm cf}\,,
\end{equation}
where
\begin{eqnarray}
\vec F_{\rm B}&=&\frac{q}{c}(\vec v_{\rm rel}\times\vec B)
{\rm\ \ is\ the\ magnetic\ force}\,,\nonumber\\
\vec F_{\rm c}&=&\left[2m\frac{dr}{dt}+r\frac{dm}{dt}\right](\hat{e}_{\rm
r}\times\vec \O )
{\rm \ \ is\ the\ Coriolis\ force,} \nonumber\\
\vec F_{\rm cf}&=&m(\vec\O\times\vec r) \times
\vec\O
{\rm\ \ is\ the\ centrifugal\ force}  \,,\nonumber
\end{eqnarray}
$c$ is the velocity of light, $q$ and $m$ are the charge and the relativistic mass of particle,
and $\vec v_{\rm rel}$ is the relative rotation velocity between particle and
magnetic field lines.
 Hence an observer in the rotating frame declares that three forces
 are acting on a particle: (i) the force $\vec F_{\rm c}$
 (Goldstein 1990) acts in the negative $\theta$-direction,
 i.e., {\it opposite} to the direction of rotation, (ii)
 the force $\vec F_{\rm B}$ which initially acts in the direction of
 $\vec v_{\rm rel}\times\vec B$, and as particle moves it turns towards the
 positive $\theta$--direction in the case of a positively charged
 particle, and (iii) the force $\vec F_{\rm cf}$ which acts radially {\it outward.}
 
    The constraint forces $\vec F_{\rm c}$ and $\vec F_{\rm B}$ act in the
direction perpendicular to the path of particle, and therefore, the work
done by them is zero. They are so strong that they
barely allow the particle to deviate even slightly from the prescribed
path. The spiral motion of particle vanishes when $\vec F_{\rm c}$ and
$\vec F_{\rm B}$ become equal and opposite.

The azimuthal--component of Eq.~(1) gives
\begin{equation}
B=\frac{c \O}{q v_{\rm rel}}\left [2m\frac{dr}{dt}+r\frac{dm}{dt}\right ]
\,.  \end{equation}
Similarly, the radial component gives
\begin{equation}
\frac{d}{dt}\left(m\frac{d r}{dt}\right)=m\O^2 r\,.
\end{equation}
Equations (2) and (3) describe the constrained  and accelerated motion
of a particle, respectively. As the particle moves radially outward,
the relative velocity between the particle and the magnetic field line
increases with respect to $r,$ and reaches maximum at the light cylinder. Since the particles
gain rotation energy as they move from the central engine to light cylinder and
centrifugal force acts along the field lines, a very strong magnetic field
is not required to drag the particles with rotating AGN. The field required
should be sufficient enough to balance the Coriolis force, as indicated by
Eq.~(2). Eventually the inertial forces overcome the tension in magnetic field lines, and
hence the field lines are swept back in the direction opposite to the sense of rotation, hence the
term toroidal twist. Beyond this point, the magnetic field is dominated by the toroidal component
$B_\phi.$ In the limit of tightly wound field lines, the magnetic tension force tends to collimate the
flow about the rotation axis (Begelman 1994).

     To find the relativistic momentum $\vec p$ of particle we solve 
Eq.~(1) numerically. Assume that at the time t=0, an electron is introduced with
an initial Lorentz factor $\g_i$ at a distance $r=r_{\rm LC}/10$ from the
rotation axis to move along the rotating magnetic field lines. The energy
of the particle is obtained using
$\eps=\sqrt{p^2 c^2+m^2_{\rm o} c^4}$ and  the Lorentz factor from
$\g=\eps/m_{\rm e} c^2.$
     
   Figure~1 shows the variation of $\g$ with respect to $r$ for the
particles with $\g_{\rm i}=10$, $50$ \& $100.$ The curves indicate  increase
of particles energy due to centrifugal force as they approach the light cylinder.
If we include radiation losses, the steepness of the curves will be
reduced.

\begin{figure}
\vskip -3.2 truecm \epsfysize=7.5 truecm \epsffile[00 350 450 790]{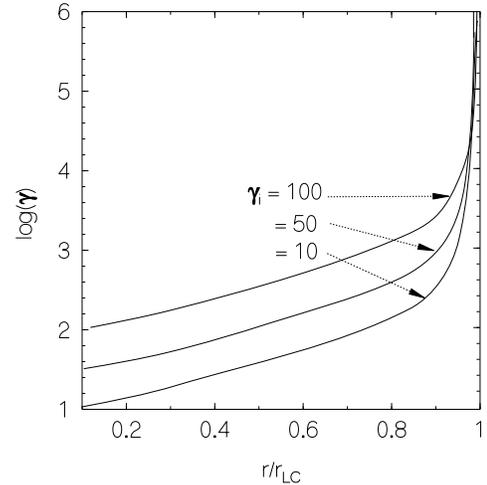}
\vskip 1.5 truecm 
   \caption{The relativistic Lorentz factor $\g$ as a
   function of $r$ at different values of $\g_{\rm i}=10$, $50$ and  $100$. }
\end{figure}

   Once the particles reach the light cylinder, they cannot return to the central engine along
the same or other field lines, as they experience a decelerating force in
their return path.  Therefore, they must leave the AGN as a jet.

Using $m=\g m_{\rm e},$ Eq.~(3) can be written as
\begin{equation}
\frac{d^2 r}{dt^2}+\e(t) \frac{d r}{dt}-\O^2 r=0\, ,
\end{equation}
where $m_e$ and $\g$ are the rest--mass and the Lorentz factor of particle,
and
       \begin{equation}
\e(t)=\frac{1}{\g}\frac{d\g}{dt}=\frac{d({\rm ln}\g)}{dt}
       \end{equation}
is a slowly increasing function of time
t for $r < r_{\rm LC}.$ Since we are interested in the behaviour of particle
trajectory and energy near light cylinder, we assume $\e(t)$ to be approximately
constant
over a small interval of time. Then the Eq.~(4) may be solved by assuming
r(t) has the form $e^{\a t},$ where $\a$ is found from
\begin{equation}
\a^2+\e \a-\O^2=0\, ,
\end{equation}
which has the solution
\begin{equation}
\a=\sqrt{\frac{\e^2}{4}+\O^2}-\frac{\e}{2}.
\end{equation}
Near the light cylinder, $\e$ takes the values much higher than $\O$ as the particles are strongly
driven in the direction of rotation,  therefore we find
\begin{equation}
\a\approx\frac{\e}{2}\left[1+2\left(\frac{\O}{\e}\right)^2\right]-
\frac{\e}{2}=\frac{\O^2}{\e}.
\end{equation}
Taking $r(0)=r_0$ as the initial condition at t=0, we obtain
\begin{equation}
r(t)=r_0 e^{\frac{\O^2}{\e} t}.
\end{equation}
Equation (9) describes the position of a particle over a small interval
of time during which $\e$ can be treated as approximately constant.

In the rest frame of an observer, let $\vec r(t)$ make an angle $\O t$ with
the x--axis. Then the co-ordinates of the position of a particle are
\begin{equation}
x(t)=r_0 e^{\frac{\O^2}{\e} t} \cos(\O t),\quad\quad
y(t)=r_0 e^{\frac{\O^2}{\e} t} \sin(\O t).
\end{equation}
Hence, the particle follows a curved trajectory as it moves along the rotating
magnetic field lines. Let $\tan\d=dy/dx$ and $ds=\sqrt{dx^2+dy^2}$
 be the slope and arc-length of the trajectory at any time t. Then
the radius of curvature of particle trajectory is
 \begin{equation}
 R=\frac{[1+(dy/dx)^2]^{3/2}}{{\vert d^2\!
 y/dx^2\vert}}
 =\sqrt{1+\left(\frac{\O}{\e}\right )^2} r(t).
\end{equation}

  Consider a typical AGN with light cylinder radius $r_{\rm LC}
=5\cdot 10^{15}$~cm and magnetic field $B=100$~G  (Lesch et al. 1989).
Using Eq.~(2), we can estimate the maximum value of $\e$ in
a magnetic field of given strength:
 \begin{equation}
\e = \frac{1}{r}\left (\frac{\o_B}{\O} v_{\rm rel}-2 \frac{dr}{dt}\right)\,
, \end{equation}
where $\o_B=q B/mc.$

	Using $v_{\rm rel}=c/100$, $dr/dt\approx c,$  $\g=10^4$ and
$r=r_{\rm LC},$ we find for an electron $\e=1.8\cdot 10^3
$~s$^{-1},$ which is much higher than $\O=6\cdot 10^{-6}$~rad~s$^{-1}.$
Therefore, Eq.~(5) gives
 \begin{equation}
\g(t)=\g_0 e^{1800\, t}\, ,
\end{equation}
where $\g_0$ is the value of $\g$ at time $t=0.$ Hence in the region near  light cylinder,
the centrifugal acceleration becomes very large and it dominates over the radiation
reaction exerted by the ambient radiation field.

A simple estimate gives us an order of magnitude estimate of the maximum
energy a particle can gain from the centrifugal acceleration.
Close to the light cylinder ($v_{\rm rel}\sim c$; $\O\sim v_{\rm rel}/r$) the minimum
time scale for acceleration is close to the inverse of the relativistic electron gyrofrequency
$\omega_B,$
\begin{equation}
t_{\rm acc}\simeq \eta^{-1}\simeq\omega_{\rm B}^{-1}={\gamma m_{\rm e} c\over {eB}}.
\end{equation}
The loss time for inverse Compton scattering is given by (e.g., Longair 1992)
\begin{equation}
t_{\rm loss}^{\rm IC}\simeq {3\cdot 10^7\over{\gamma U_{\rm rad}}}\, {\rm s}
\end {equation}
and for synchrotron losses
\begin{equation}
t_{\rm loss}^{\rm SYN}\simeq {5\cdot 10^8\over{\gamma B^2}}\, {\rm s},
\end {equation}
where $U_{\rm rad}$ denotes the energy density of the incoming radiation. We can estimate
$U_{\rm rad}$ of the central luminosity $L$ at a distance $R$ as
$ U_{\rm rad}\simeq {L/{4\pi R^2 c}}.$
Using the definition of non-relativistic electron gyrofrequency
$\omega_{\rm ce}=eB/{m_{\rm e} c},$
we obtain the expression for maximum Lorentz factor an electron can achieve via
centrifugal acceleration, including inverse Compton scattering and/or synchrotron
radiation,
\begin{equation}
\gamma^{\rm IC}_{\rm max}\simeq \sqrt{{3\cdot 10^7 \omega_{\rm ce}\over{U_{\rm rad}}}}
{\quad \ \ {\rm and}\quad }
\gamma^{\rm SYN}_{\rm max}\simeq \sqrt{{5\cdot 10^8 \omega_{\rm ce}\over{B^2}}}.
\end{equation}
The ratio of both is
\begin{equation}
{\gamma_{\rm max}^{\rm SYN}\over{\gamma_{\rm max}^{\rm IC}}}=
\sqrt{8\pi}{\sqrt{U_{\rm rad}}\over{B}}.
\end{equation}
Hence depending upon the radiation intensity and the field strength, one of the
radiation mechanisms will limit the centrifugal acceleration.

         Figure~2 shows both the Lorentz factors for a central luminosity of
$L=10^{46}$~{\rm erg s}$^{-1}$ over the central 100 Schwarzschild-radii of a black hole
with $10^8\, M_\odot$. The toroidal component of magnetic field is assumed 
scale as $B(R)\propto B_0 R^{-1}$. With $B_0=1$~kG we still
have $U_{\rm rad}>B(R)^2/8\pi$ (Fig.~3),  which means inverse Compton scattering is faster
than synchrotron radiation, and that is the reason why $\gamma_{\rm max}^{\rm SYN}\gg
\gamma_{\rm max}^{\rm IC}.$

\begin{figure}
\vskip -3.0 truecm \epsfysize=8.0 truecm \epsffile[00 350 450 790]{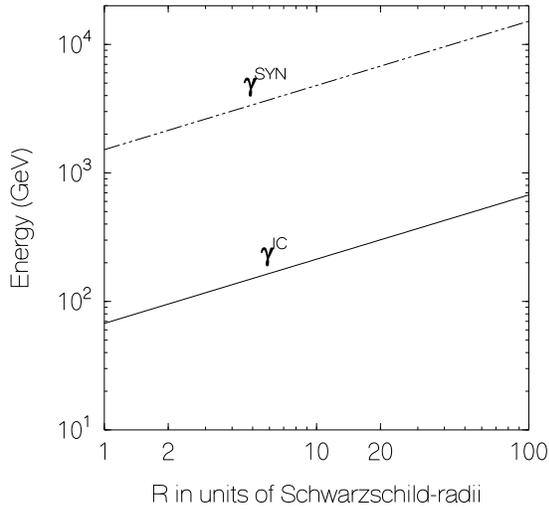}
\vskip 1.7 truecm
\caption{The relativistic Lorentz factors $\g^{\rm IC}$ and
$\g^{\rm SYN}$ as functions of $R.$}
\end{figure}

\begin{figure}
\vskip -3.0 truecm \epsfysize=8.0 truecm \epsffile[00 350 450 790]{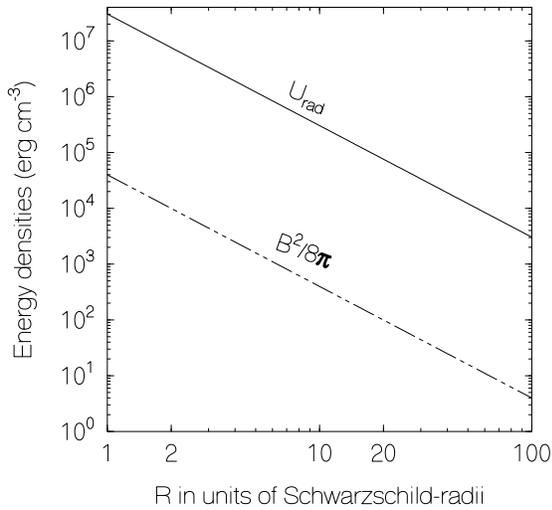}
\vskip 1.7 truecm
\caption{The radiation energy density and magnetic energy density
as functions of $R.$}
\end{figure}

         Our results imply that continuously a population of 10~GeV electrons escape
from the central $10^{15}$~cm region (the light cylinder), in which they scatter the UV-photons
to the X--ray and/ $\gamma$--ray ranges (von Montigny et~al. 1995). Outside the 
light cylinder,
the particles may encounter weaker magnetic fields (e.g. Blandford 1990) along which they can easily
generate the observed steady nonthermal optical emission of AGN via synchrotron
radiation:
\begin{equation}
\nu\simeq 10^{14} Hz \left[{\gamma\over{5000}}\right]^2\left[{B\over{1G}}\right]
\end{equation}

\section {Conclusion}
A simple approximation to the plasma dynamics, is the study of motion of individual
plasma particles, and it is a good approximation for the rarefied plasma.
We consider a physical mechanism which relates the MHD-scenarios for
the production of relativistic jets via rotating magnetospheres (e.g. Camenzind 1996)
and particle acceleration, visible as continuous nonthermal
radiation ranging from gamma-radiation down to optical and radio emission.
The charged particles moving along the rotating magnetic field lines experience the centrifugal
acceleration.  Depending upon the stiffness of the magnetic field lines, this can
become very strong compared to the inverse Compton scattering and
synchrotron radiation losses. Our model provides  a natural explanation for an efficient way of
extracting the rotational energy from central engines and the continuous escape of
the high energy electrons from AGN into the jet material. They  scatter the
UV--photons to the X--ray and $\gamma$--ray ranges at the light cylinder, and outside they generate
the steady nonthermal optical emission of AGN via synchrotron radiation. It may
even account for the continuous radio emission, taking into account that
the particles loose energy (i.e. $\gamma$ drops) and the magnetic field strengths
decreases as well (Blandford 1990).
\begin{acknowledgements}
 RTG is thankful to Prof. R. Wielebinski for inviting to the Max--Planck Institut 
f\"ur Radioastronomie. This work was supported by the Alexander von Humboldt 
Fellowship.
 \end{acknowledgements}

\end{document}